# A Route Toward the On-Surface Synthesis of Organic Ferromagnetic Quantum Spin Chains

Fabian Paschke[1,*], Ricardo Ortiz[2], Shantanu Mishra[1,*], Manuel Vilas-Varela[3], Florian Albrecht[1], Diego Peña[3,4,*], Manuel Melle-Franco[2], Leo Gross[1,*]

[1]IBM Research Europe – Zurich, 8803 Rüschlikon, Switzerland
[2]CICECO - Institute of Materials, Department of Chemistry, University of Aveiro, Aveiro 3810-193, Portugal
[3]Center for Research in Biological Chemistry and Molecular Materials (CiQUS) and Department of Organic Chemistry, University of Santiago de Compostela, 15702 Santiago de Compostela, Spain
[4]Oportunius, Galician Innovation Agency (GAIN), 15702 Santiago de Compostela, Spain

[*]Corresponding authors: FAP@zurich.ibm.com, SHM@zurich.ibm.com, diego.pena@usc.es and LGR@zurich.ibm.com

## Abstract

**Engineering sublattice imbalance is an intuitive way to induce high-spin ground states in bipartite polycyclic conjugated hydrocarbons. Such high-spin molecules can be employed as building blocks of quantum spin chains, which are outstanding platforms to study many-body physics and fundamental models in quantum magnetism. This is exemplified by recent reports on the bottom-up synthesis of antiferromagnetic molecular spin chains that provided insights into paradigmatic quantum phenomena such as fractionalization. In contrast to antiferromagnetism, demonstration of ferromagnetic coupling between polycyclic conjugated hydrocarbons has been scarce. Previous attempts in this direction were limited by the formation of non-benzenoid rings leading to spin quenching, or the use of spacer motifs that considerably weaken the magnitude of ferromagnetic exchange. Here, we demonstrate the on-surface synthesis of short ferromagnetic spin chains based on dibenzotriangulene, a polycyclic conjugated hydrocarbon with a triplet ground state. Our synthetic strategy centers on the simple concept of achieving a direct (that is, without a spacer motif) majority-minority sublattice coupling between adjacent units. This leads to a global sublattice imbalance in spin chains scaling with the chain length, and therefore a ferromagnetic ground state with a strong intermolecular ferromagnetic exchange. By means of scanning probe measurements and multiconfigurational quantum chemistry calculations, we analyze the electronic and magnetic properties of ferromagnetic dimers and trimers of dibenzotriangulene, and confirm their quintet and septet ground states, respectively, with an intermolecular ferromagnetic exchange of 7 meV. Furthermore, we elucidate the role of sublattice coupling on magnetism through complementary experiments on antiferromagnetic dibenzotriangulene dimers with majority-majority and minority-minority couplings. We expect our proof-of-principle study to provide impetus for the design of purely organic ferromagnetic materials.**

The study of organic radicals, in particular open-shell polycyclic conjugated hydrocarbons (PCHs), is an active research area fueled by advances in on-surface chemistry and scanning probe techniques that allow synthesis, stabilization and atomic-scale characterization of reactive species on surfaces. The interest in open-shell PCHs stems from the fundamental insights they provide into quantum magnetism along with potential application in optoelectronic and spintronic technologies.[1] A typical example of open-shell PCHs is the family of [$n$]triangulenes (Fig. 1a) for which no Kekulé valence structures can be drawn without leaving at least $n - 1$ unpaired electrons ($n$ denotes the number of benzenoid rings along an edge). The magnetic ground state of bipartite PCHs can be intuitively predicted by Ovchinnikov's rule,[2] which states that the ground-state total spin quantum number $S = |N_A - N_B|/2$, where $N_A$ and $N_B$ denote the number of carbon atoms in the two interpenetrating sublattices of the underlying bipartite lattice (Fig. 1a). For [$n$]triangulenes, $|N_A - N_B| = n - 1 > 0$, leading to a high-spin ($S > 0$) ground state with a linear scaling of $S$ with $n$. Derivatives of [3]triangulene (hereafter, triangulene) have been obtained in solution,[3,4] while unsubstituted [$n$]triangulenes with $n$ between 2 and 7 have been synthesized on metallic and insulating surfaces.[5–9]

In contrast to the localized nature of magnetic moments in molecules and solids containing transition metal ions, unpaired electrons in open-shell PCHs are hosted in delocalized molecular orbitals that extend over several carbon atoms. This opens avenues for engineering robust and tunable intermolecular magnetic exchange interactions in covalently bonded nanostructures of open-shell PCHs. For instance, Fig. 1b shows a triangulene dimer, where the individual triangulene units are connected *via* a single carbon-carbon bond through their minority sublattice atoms. In this case, there is no global sublattice imbalance in the dimer ($N_A - N_B = 0$), and Ovchinnikov's rule predicts a singlet ($S = 0$) ground state, that is, an antiferromagnetic exchange between the triangulene units. Introduction of a 1,4-phenylene spacer between the triangulene units (Fig. 1b) maintains an antiferromagnetic exchange. However, it weakens the magnitude of intermolecular exchange interaction primarily because of a reduced intermolecular hybridization of the frontier molecular orbitals.[10] The structures shown in Fig. 1b were experimentally realized, and proven to exhibit an antiferromagnetic ground state with the exchange interactions roughly being 2 and 14 meV for dimers with and without the spacer group, respectively.[11] Building up on these results, spectacular progress has been made in recent years toward the bottom-up synthesis of antiferromagnetic spin chains of [n]triangulenes and related molecules, which have been used as a platform to probe fundamental physics of antiferromagnetic quantum spin chains such as fractionalization and the valence-bond solid state.[12–16]

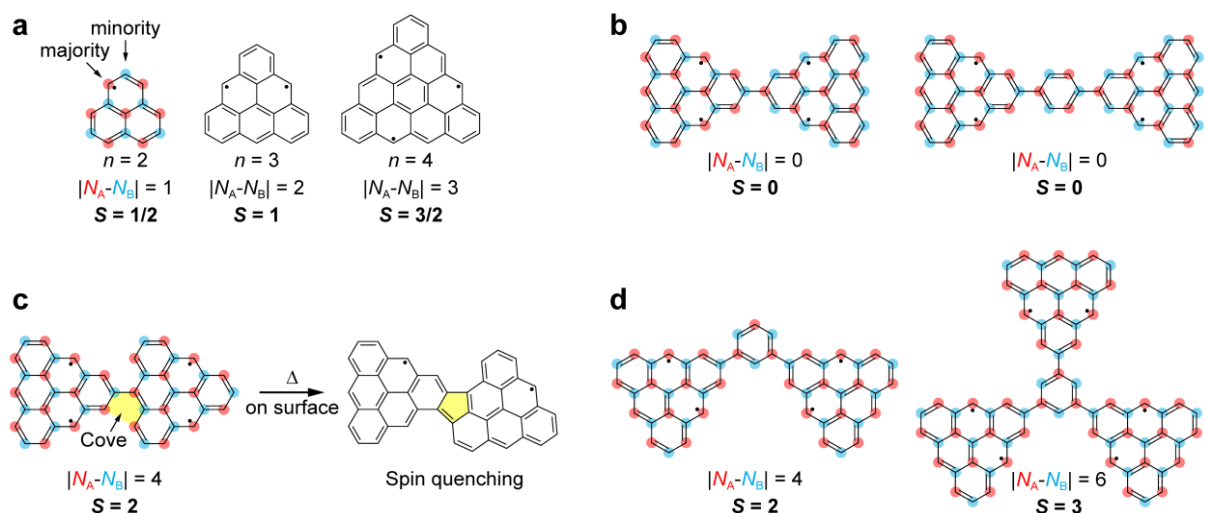

**Figure 1. Concepts in generation of ferromagnetic and antiferromagnetic exchange in triangulene dimers and trimers. (a)** Chemical structure of [n]triangulenes with carbon atoms of the two sublattices highlighted in red ($N_A$) and blue ($N_B$). Majority and minority sublattices are highlighted for [2]triangulene. **(b)** Antiferromagnetic triangulene dimers with minority-minority coupling (left) and coupling *via* a 1,4-phenylene spacer (right). Here, both sublattices contain an equal number of carbon atoms. **(c)** Ferromagnetic triangulene dimer with majority-minority coupling highlighting the formation of a pentagonal ring at the cove region that leads to spin quenching. **(d)** Experimentally realized ferromagnetic triangulene dimer and trimer obtained by coupling triangulene units *via* a 1,3-phenylene (left) and a 1,3,5-phenylene (right) spacer.

In contrast to antiferromagnetism, progress on the design and characterization of ferromagnetically coupled open-shell PCHs has been relatively scarce.[17–20] From Ovchinnikov's rule, ferromagnetic coupling is obtained when there is a global sublattice imbalance ($|N_A - N_B| > 0$). For a triangulene dimer, the conceptually simplest route to achieve a ferromagnetic coupling is to directly connect the majority sublattice of one triangulene unit with the minority sublattice of the other, that is, a majority-minority coupling (Fig. 1c). Note that minority-minority (Fig. 1b) and majority-majority couplings do not lead to a global sublattice imbalance. However, the problem with this approach is that it generates cove regions that invariably leads to the cyclodehydrogenative formation of pentagonal rings between the triangulene units during thermal annealing at temperatures necessary to achieve on-surface synthesis of the dimers (Fig. 1c). The presence of a pentagonal ring breaks the bipartite symmetry of the lattice and leads to spin quenching, as has been experimentally observed in two recent works.[21,22] A route to avoid this problem is to introduce spacer groups between the triangulene units. In contrast to a 1,4-phenylene spacer that does not lead to a global sublattice imbalance (Fig. 1b), a 1,3-phenylene spacer results in a global sublattice imbalance, and therefore a ferromagnetic exchange between the triangulene units (Fig. 1d). Two recent works have reported the generation of triangulene dimers and trimers with phenyl and triazinyl spacer groups to achieve ferromagnetic exchange between the constituent triangulene units (Fig. 1d).[18,20] However, this approach leads to a weak intermolecular exchange of about 1 meV, due to the increased spatial separation between the triangulene units.[10]

Here, we demonstrate a route to achieve ferromagnetic spin chains based on dibenzotriangulene (DBT, Fig. 2a), a dibenzo-extended triangulene derivative that has $|N_A - N_B| = 2$ and therefore a triplet ($S = 1$) ground state. Our approach is based on the design and solution synthesis of the precursor 2,9-dibromo-6-(2,6-dimethylphenyl)pentacene (**1**, Fig. 2b), which undergoes dehalogenative aryl-aryl coupling and subsequent oxidative cyclization upon thermal annealing on a Au(111) surface (Fig. 2b–e). Substitution of the pentacene core of **1** by Br atoms at the 2,9 positions enables a direct majority-minority coupling of the constituent DBT units that leads to a strong ferromagnetic exchange. Importantly, the benzo extension leads to gulf regions (Fig. 2d) that prevents pentagonal ring formation, thereby circumventing the spin quenching observed in previous approaches relying on direct majority-minority coupling of triangulene units. By means of high-resolution scanning tunneling microscopy (STM), scanning tunneling spectroscopy (STS) and non-contact atomic force microscopy (AFM), we perform structural and electronic characterization of ferromagnetic dimers and trimers of DBT as model systems for ferromagnetic quantum spin chains. Compound **1** also affords the formation of antiferromagnetic dimers with minority-minority and majority-majority couplings (Fig. 2c and e), whose electronic structure we compare and contrast with the ferromagnetic dimer. Our experimental results are supported by mean-field and multiconfigurational quantum chemistry calculations.

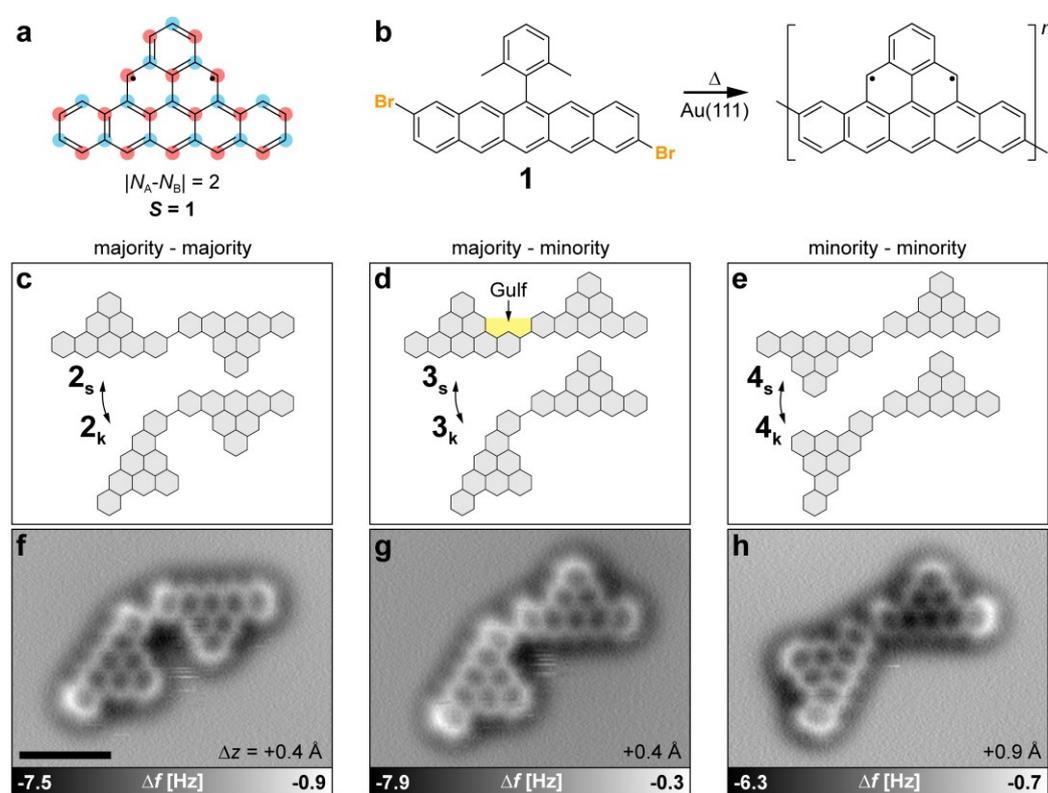

**Figure 2. On-surface synthesis of DBT dimers on Au(111). (a)** Chemical structure and sublattice representation of DBT. **(b)** On-surface synthetic route to DBT spin chains. **(c–e)** Schematic representation of the three DBT dimer classes based on the relative sublattice coupling of the constituent DBT units. Double arrows indicate the two possible conformations by rotation of one unit relative to the other on the surface. The subscripts s and k denote the straight and kinked conformation, respectively. The gulf region that prevents formation of a pentagonal ring is highlighted. **(f–h)** Corresponding AFM images of kinked DBT dimers on Au(111). Scale bar: 1 nm (applies to panels f–h).

## Results/Discussion

**On-surface synthesis of dibenzotriangulene dimers**

The fabrication of DBT spin chains is based on the solution synthesis of precursor **1** (Figs. S1–S3). A submonolayer coverage of **1** was deposited on a Au(111) surface and annealed up to 360 °C to promote surface-catalyzed dehalogenative aryl-aryl coupling and subsequent oxidative cyclization of the methyl groups (Fig. 2b, see also Methods). Annealing resulted in a mixture of individual DBT molecules, dimers and longer oligomers on the surface (Figs. S4 and S5). We first focus on the structural and electronic characterization of DBT dimers as model systems for spin chains, followed by a discussion of trimers later in the article. The 2,9-brominated sites in **1** afford three different coupling motifs for a dimer, depending on whether the carbon-carbon bond between two DBT units connects atoms that belong to the majority or minority sublattices of the respective DBT units. Therefore, DBT dimers can be categorized in three classes: dimers that exhibit majority-majority coupling (**2**, Fig. 2c), majority-minority coupling (**3**, Fig. 2d) and minority-minority coupling (**4**, Fig. 2e). Furthermore, for each class, the dimers can adopt a straight (**2$_s$**, **3$_s$** and **4$_s$**) or a kinked (**2$_k$**, **3$_k$** and **4$_k$**) geometry on the surface, owing to the prochirality of the DBT moieties. Experimentally, dimers with all of these sublattice couplings and geometries were found on the surface (Fig. S6). The electronic structure (namely, transport gaps and spin excitation energies) of dimers of a given class are independent of their geometry, as was shown in a previous work on triangulene spin chains.[12] We focus on kinked DBT dimers in the main text (see Fig. 2f–h for the corresponding AFM images) and provide results on straight dimers in Fig. S7.

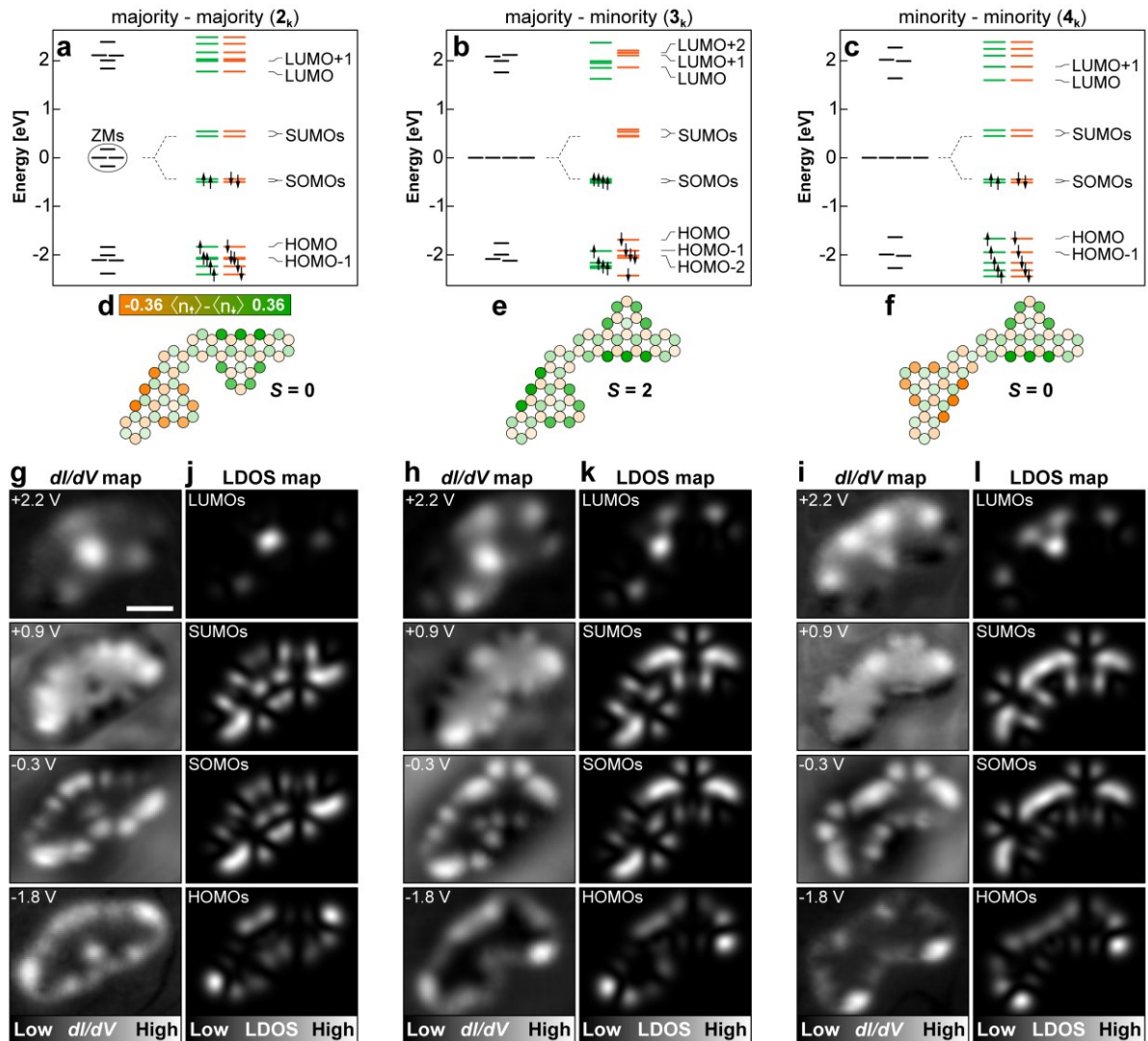

**Figure 3. Electronic structure of DBT dimers.** **(a–c)** Tight-binding (black markers) and MFH (colored markers) ground-state energy spectra of **2$_k$**, **3$_k$** and **4$_k$**. Green (orange) markers denote spin up (down) levels. **(d–f)** MFH spin polarization plots of **2$_k$**, **3$_k$** and **4$_k$** in their respective ground states. Green (orange) filled circles denote mean populations of spin up (down) electrons. The color scale applies to panels d–f. **(g–i)** $dI/dV$ maps of the DBT dimers whose AFM images are shown in Fig. 2f–h with bias voltages indicated in the respective panels. STM parameters: $I$ = 300 pA; root mean squared modulation voltage $V_{rms}$ = 50 mV. **(j–l)** Corresponding MFH-LDOS maps of the HOMOs to LUMOs of **2$_k$**, **3$_k$** and **4$_k$** (see text for details). Scale bar: 1 nm (applies to all figures in panels g–l).

**Electronic characterization of dibenzotriangulene dimers**

We start by exploring the electronic structure of an individual DBT molecule and DBT dimers within the nearest-neighbor tight-binding model. As shown in Fig. S5, the key feature in the tight-binding spectrum of DBT is the presence of two zero-energy modes (ZMs), corresponding to two unpaired electrons, which are localized on the majority sublattice atoms. The ZMs of individual DBT units survive upon formation of the dimers (Fig. 3a–c), leading to four ZMs in each dimer. In **2$_k$**, coupling of the individual DBT units *via* their majority sublattice atoms results in a non-negligible hybridization between one of the two ZMs from each DBT unit (since the sites where these ZMs are localized are nearest neighbors), which manifests as a small gap. The other pair of ZMs do not hybridize and thus remain degenerate and at zero energy.

We next analyze the systems in the mean-field Hubbard (MFH) approximation to describe the formation of local magnetic moments and magnetic exchange (Fig. 3a–c). For DBT, the degeneracy of the ZMs is lifted in the MFH approximation, leading to the formation of two singly occupied molecular orbitals (SOMOs) and the associated unoccupied molecular orbitals (SUMOs), with a triplet ground state, that is, a ferromagnetic exchange between the unpaired electrons in the SOMOs (Fig. S5). For the dimers **2$_k$** and **4$_k$**, the lowest energy MFH solution corresponds to an antiferromagnetic exchange between the constituent DBT units, that is, a singlet ground state (Fig. 3a, c, d and f), while for **3$_k$**, the ground state is a quintet ($S$ = 2), implying a ferromagnetic exchange between the DBT units (Fig. 3b and e). The results of our MFH calculations agree with Ovchinnikov's rule: **2$_k$** and **4$_k$** have no sublattice imbalance which results in a singlet ground state, while **3$_k$** exhibits a global sublattice imbalance ($|N_A – N_B|$ = 4), resulting in a quintet ground state.

Experimentally, we accessed the electronic structure of **2$_k$**, **3$_k$** and **4$_k$** by recording differential conductance ($dI/dV$, $I$ and $V$ correspond to the tunneling current and bias voltage, respectively) maps of the dimers (Fig. 3g–i) and comparing them to the MFH local density of states (LDOS) maps of the molecular orbitals of the dimers (Fig. 3j–l). From the $dI/dV$ maps, we identify four distinct molecular orbital resonances between $V$ = -2.2 V and +2.2 V, that is, at $V$ = -1.8, -0.3, +0.9 and +2.2 V. The $dI/dV$ maps at -0.3 and +0.9 V exhibit similar shapes and symmetries for each dimer, which strongly suggests that they are associated with tunneling from (into) the SOMOs (SUMOs).[5,23] This is supported by the good agreement of the $dI/dV$ maps at -0.3 and +0.9 V with the MFH-LDOS maps of the SOMOs and SUMOs of the dimers, respectively (Fig. 3g–l). On the other hand, for each dimer, the $dI/dV$ maps at -1.8 and +2.2 V show pronounced differences, suggesting that they are associated with tunneling through hybridized molecular orbitals. For **2$_k$** and **4$_k$**, the $dI/dV$ maps at -1.8 and +2.2 V are best reproduced in MFH-LDOS maps by considering equal contributions of HOMO/HOMO-1 and LUMO/LUMO+1, respectively. For **3$_k$**, the $dI/dV$ maps at -1.8 and +2.2 V are well reproduced by considering equal contributions of HOMO/HOMO-1/HOMO-2 and LUMO/LUMO+1/LUMO+2, respectively.

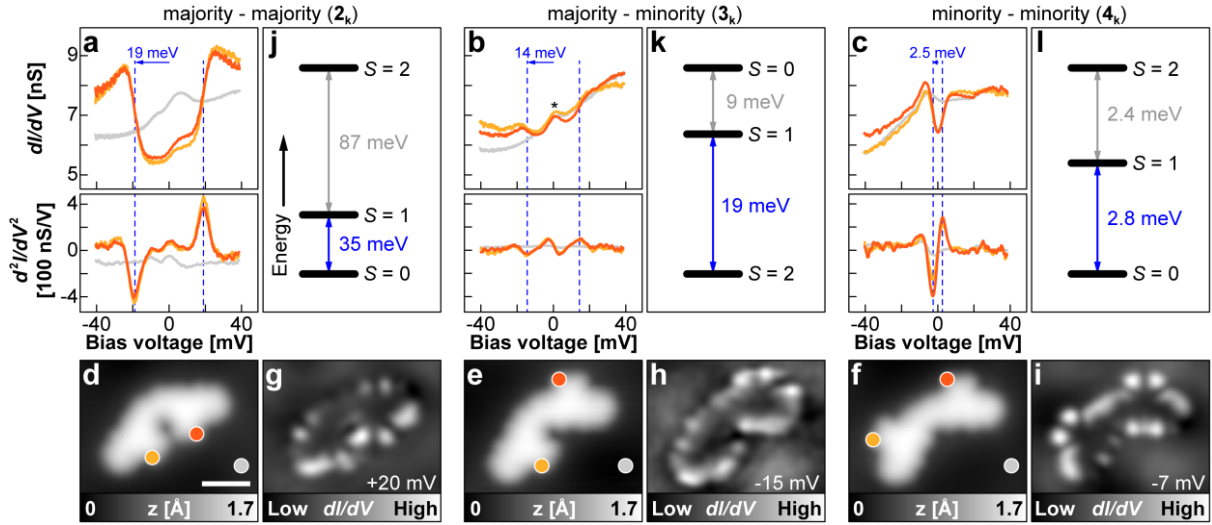

**Figure 4. Magnetic properties of DBT dimers.** **(a–c)** $dI/dV(V)$ (top) and $d^2I/dV^2(V)$ (bottom) spectra obtained on **2$_k$**, **3$_k$** and **4$_k$**. Background spectra on Au(111) are shown in grey. Dashed lines denote extracted spin excitation energies (see Methods). The asterisk in panel b denotes the $S$ = 2 Kondo resonance of **3$_k$**. STM open-feedback parameters: $V$ = +40 mV, $I$ = 300 pA; $V_{rms}$ = 3.5 mV. **(d–f)** STM images of the DBT dimers. Colored circles indicate the positions at which spectra in panels a–c have been recorded. STM parameters: $V$ = +40 mV, $I$ = 300 pA. **(g–i)** $dI/dV$ maps of the DBT dimers obtained around the respective spin excitation thresholds as indicated. STM parameters: $I$ = 300 pA; $V_{rms}$ = 3.5 mV. **(j–l)** Calculated CASSCF-NEVPT2 energies of the spin ground and excited states of the DBT dimers. Blue arrows denote the energy differences between the ground and first excited states. Scale bar: 1 nm (applies to panels d–i).

**Magnetic excitations in dibenzotriangulene dimers**

Figure 4a–f shows $dI/dV(V)$ spectra acquired on DBT dimers within a small bias window of ±40 mV, where the salient features are abrupt steps, symmetric around zero bias. We ascribe these spectral features to inelastic spin excitations, and their energies indicate the energetic differences between the magnetic ground state and the excited states. It is noted that spin excitations observed by STS follow the spin selection rule[24] that the change in total spin quantum number $\Delta S$ must be 0 or ±1. The spin excitation energies, extracted from the corresponding experimental $d^2I/dV^2(V)$ signal, are 19 meV ($2_k$), 14 meV ($3_k$) and 2.5 meV ($4_k$) (Fig. 4a–c). Note that the $dI/dV$ maps acquired at bias voltages near the excitation thresholds for all three dimers resemble the orbital densities of their respective SOMOs, which supports the magnetic origin of these excitations (Fig. 4g–i).

To relate the observed spin excitation energies to transitions between the ground and excited states of the dimers, we performed multi-reference many-body perturbation theory calculations. Briefly, we applied second-order n-electron valence state perturbation theory (NEVPT2) to complete active space self-consistent field (CASSCF) wavefunctions, with the choice of active space being 8 electrons in 12 single-particle states, that is, CASSCF(8,12) (see Methods). The CASSCF-NEVPT2 calculations confirm the singlet ground states of $2_k$ and $4_k$ as well as the quintet ground state of $3_k$ (Fig. 4j–l). The energetic order of spin states for the antiferromagnetic dimers $2_k$ and $4_k$ is singlet < triplet < quintet. Together with the spin selection rule, this implies that the observed spin excitations at 19 and 2.5 meV for $2_k$ and $4_k$, respectively, correspond to singlet-triplet excitations. The energetic order of spin states for the ferromagnetic dimer $3_k$ is quintet < triplet < singlet, and therefore the observed spin excitation at 14 meV is assigned to a quintet-triplet excitation.

Based on our experiments and calculations on the DBT dimers, we draw the following conclusions:

**(1)** Comparing the two antiferromagnetic dimers $2_k$ and $4_k$, their experimental (calculated) singlet-triplet excitation energies differ by an order of magnitude, that is, 19 meV (35 meV) for $2_k$ compared to 2.5 meV (2.8 meV) for $4_k$. We rationalize this difference as follows: for $2_k$ and $4_k$, the dominant exchange mechanism is the intermolecular kinetic superexchange that is driven by the intermolecular hybridization between the ZMs of individual DBT units.[10] The magnitude of kinetic superexchange scales quadratically with the intermolecular hybridization. While the ZMs of the two DBT units in $2_k$ are nearest neighbors (and thus hybridize strongly), the ZMs of the DBT units in $4_k$ are third-nearest neighbors, leading to a much weaker hybridization driven by third-nearest neighbor hopping in the tight-binding framework.

**(2)** For the ferromagnetic dimer $3_k$, in addition to the quintet-triplet spin excitation, we also observe a small peak at zero bias (denoted with an asterisk in Fig. 4b). Given the high-spin ground state of $3_k$, we associate this peak with the Kondo resonance of an $S = 2$ ground state, as has been recently observed in a similar system.[20]

**(3)** Given the strong intra-DBT ferromagnetic exchange (roughly 200 meV from MFH calculations, see Fig. S5) and weaker inter-DBT exchange, along with extremely weak magnetic anisotropy in carbon-based systems, it seems reasonable to consider $2_k$, $3_k$ and $4_k$ as weakly-coupled $S = 1$ Heisenberg dimers whose energy spectrum is described by the Hamiltonian $\hat{H} = J_{\text{eff}} \hat{S}_1 \cdot \hat{S}_2$, where $J_{\text{eff}}$ denotes the effective intermolecular exchange ($J_{\text{eff}} < 0$ for $3_k$, while $J_{\text{eff}} > 0$ for $2_k$ and $4_k$), and $\hat{S}_1$ and $\hat{S}_2$ denote $S = 1$ operators at the two DBT sites. Should this approximation hold true, the energetic difference between the spin states for the dimers, irrespective of their ground state, should be $E_{\text{triplet}} - E_{\text{singlet}} = J_{\text{eff}}$, and $E_{\text{quintet}} - E_{\text{triplet}} = 2J_{\text{eff}}$.[11] From the many-body calculations in Fig. 4j–l it is seen that $3_k$ can be well approximated by the Heisenberg dimer model, based on which we extract the experimental $J_{\text{eff}} = 7$ meV. For $2_k$ and $4_k$, the Heisenberg dimer model does not agree well with the CASSCF-NEVPT2 results. This can be rationalized by the fact that in such antiferromagnetic dimers, additional non-linear exchange mechanisms, which are higher-order manifestations of the intermolecular kinetic superexchange, may cause deviations from a pure Heisenberg dimer model.[12,25]

**(4)** Finally, we note that discrepancies between the experimental and CASSCF-NEVPT2 excitation energies likely arise because of the effect of the underlying surface, which is not considered in our calculations. It has been recently shown in a theoretical work that compared to gas phase, adsorption of open-shell PCHs on metallic surfaces causes renormalization of the spin excitation energies.[26]

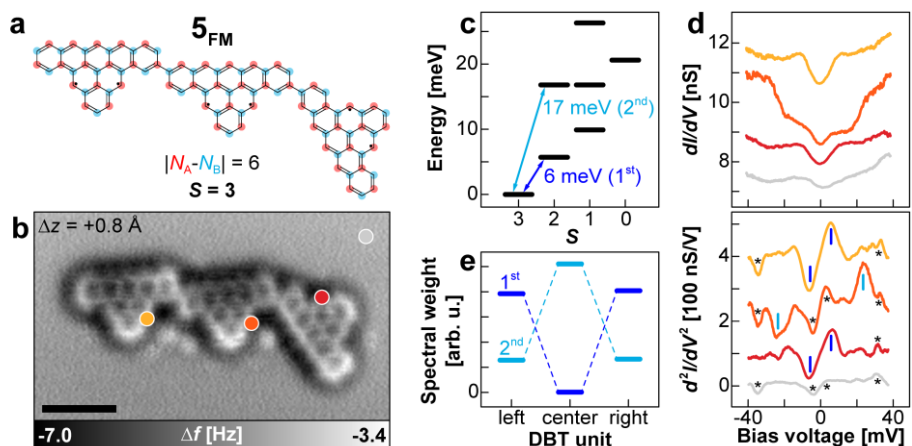

**Figure 5. Magnetic properties of a ferromagnetic DBT trimer. (a)** Chemical structure and sublattice representation of **5$_{FM}$**. **(b)** Corresponding AFM image of **5$_{FM}$**. Colored circles indicate positions at which spectra shown in panel d have been recorded. **(c)** Calculated CASSCF-NEVPT2 energies of the spin ground and excited states of **5$_{FM}$**. Arrows denote the energy differences between the (septet) ground and (quintet) excited states that obey the spin selection rule. **(d)** $dI/dV(V)$ (top) and $d^2I/dV^2(V)$ (bottom) spectra obtained on **5$_{FM}$** using a CO tip. A background spectrum obtained on Au(111) is shown in grey. Asterisks denote the vibrational modes of CO and solid blue lines denote extracted spin excitation energies. For clarity, $dI/dV$ ($d^2I/dV^2$) spectra on left, central and right units are vertically shifted by 4 nS (400 nS/V), 2 nS (250 nS/V) and 1 nS (100 nS/V), respectively. STM open-feedback parameters: $V$ = +40 mV, $I$ = 300 pA; $V_{rms}$ = 3.5 mV. **(e)** Spatially resolved CAS spectral weights of the two lowest septet-quintet excitations at 6 meV and 17 meV (denoted 1$^{st}$ and 2$^{nd}$ in panel c). Scale bar: 1 nm.

**Magnetic excitations in ferromagnetic dibenzotriangulene trimers**

Figure 5a and b display the chemical structure and AFM image of a ferromagnetic DBT trimer (**5$_{FM}$**), respectively, in which the constituent DBT units are connected *via* majority-minority coupling, leading to a septet ($S = 3$) ground state as per Ovchinnikov's rule. CASSCF-NEVPT2 calculations confirm a septet ground state of **5$_{FM}$**, which is followed by a quintet (6 meV), a triplet (10 meV) and another quintet (17 meV) state (Fig. 5c). $dI/dV(V)$ spectra acquired on **5$_{FM}$** reveal spin excitations at 6 meV on the outer two DBT units and 23 meV on the central unit (Fig. 5d). Note that the excitations at around 3 and 32 meV (denoted with asterisks in Fig. 5d) are vibrational excitations corresponding to the frustrated translational and rotational modes, respectively, of the carbon monoxide (CO) molecule at the tip apex (see Fig. S8 for measurements with a metallic tip, where these excitations are absent).[27] Based on our CASSCF-NEVPT2 results and the spin selection rule, the observed spin excitations at 6 and 23 meV are assigned to septet-quintet excitations. The presence or absence of the two septet-quintet excitations on the outer or central DBT units can be linked to the spectral weights of the corresponding excitations,[12,24] which determine the probability of excitation to the quintet states through spin-dependent electron tunneling across a given DBT unit. Figure 5e shows the CAS spectral weights of the two septet-quintet excitations (see Methods for details). The lower energy excitation at 6 meV (denoted as 1$^{st}$) has maximum weight on the outer DBT units with no weight on the central unit, whereas the higher energy excitation at 17 meV (denoted as 2$^{nd}$) has maximum weight on the central DBT unit and a very small weight on the outer units, in line with the experiments.

## Conclusions

By means of on-surface synthesis we demonstrated the regioselective coupling of triplet DBT units into dimers and trimers. Per unit, each sublattice hosts one binding site, enabling three possible intermolecular bonding motifs. By comparing spin excitation energies of the dimers to many-body calculations, we verified antiferromagnetic exchange for majority-majority and minority-minority couplings, and a ferromagnetic exchange for majority-minority coupling. For the ferromagnetic DBT dimer we determined an experimental exchange of 7 meV, which is considerably larger than the exchange in ferromagnetic dimers based on triangulene units with spacer motifs (< 1 meV).[20] Our design principle of constructing spin chains based on DBT therefore opens a route to synthesize organic quantum spin chains with robust ferromagnetic coupling. We demonstrated first steps in this direction by characterization of a hexaradical ferromagnetic DBT trimer that exhibits a septet ground state.

## Methods/Experimental

**Sample preparation and STM/AFM experiments**

Measurements were performed with a home-built combined STM/AFM setup operated at a temperature of 5 K and base pressure below $1 \times 10^{-10}$ mbar.[28] The Au(111) single crystal surface was prepared by repeated cycles of sputtering with Ne$^+$ ions and annealing up to 820 K. Submonolayer coverage of **1** on the surface was obtained by flashing an oxidized silicon wafer containing precursor molecules (**1**) in front of the cold sample in the microscope. To induce polymerization *via* dehalogenative aryl-aryl coupling, the sample temperature was ramped to 493 K within 5 minutes and maintained for 3 minutes. Subsequently, the temperature was ramped to 633 K in < 1 minute and maintained for 3 minutes to induce oxidative cyclization. CO molecules for tip functionalization were deposited onto the cold sample surface (maximum sample temperature 12 K).

Unless noted otherwise, all STM and $dI/dV$ data were acquired with metallic PtIr tips covered with gold by repeated indentation into the sample surface. STM images were recorded in the constant-current mode. The bias voltage $V$ was applied to the sample with respect to the tip. $dI/dV(V)$ and $d^2I/dV^2(V)$ spectra were recorded at a fixed tip height with the feedback opened at the respective acquisition positions. Spatially resolved $dI/dV$ maps were recorded in the constant-current mode. All $dI/dV$ measurements were performed using a lock-in amplifier, modulating the bias voltage with a frequency of 163 Hz. The modulation amplitudes (root mean squared values) are given in the respective figure captions. Spin excitation energies were extracted from the data by fitting the respective dips (at $V < 0$) and peaks (at $V > 0$) in the $d^2I/dV^2(V)$ spectra using Lorentzian curves.

AFM measurements were performed in the constant-height mode using a qPlus sensor[29] operating in the frequency-modulation mode,[30] with the oscillation amplitude kept constant at 0.5 Å. AFM data were obtained by using CO-functionalized tips[31,32] and at $V = 0$ V. Positive tip-height offsets $\Delta z$ denote tip retraction after switching off the STM feedback loop at a setpoint of $V = +10$ mV and $I = 10$ pA on Au(111).

**Tight-binding and mean-field Hubbard calculations**

Calculations in the tight-binding and MFH approximation were performed by solving the mean-field Hubbard Hamiltonian

$$\hat{H} = \sum_\sigma^{\uparrow,\downarrow} \sum_{i,j} t_{ij}\left(a_{i,\sigma}^\dagger a_{j,\sigma} + c.c.\right) + U \sum_i \left(a_{i,\uparrow}^\dagger a_{j,\uparrow}\langle n_{i,\downarrow}\rangle + a_{i,\downarrow}^\dagger a_{j,\downarrow}\langle n_{i,\uparrow}\rangle - \langle n_{i,\downarrow}\rangle\langle n_{i,\uparrow}\rangle\right),$$

where $a_{i,\sigma}^\dagger$ and $a_{i,\sigma}$ denote the electron creation and annihilation operators at site $i$ with a spin $\sigma \in \{\uparrow,\downarrow\}$. The first term is the TB Hamiltonian, in which $t_{ij}$ denotes the hopping energy between site $i$ and $j$. Here, we consider nearest-neighbor hopping only ($t_1 = -2.8$ eV). The second term is the MFH Hamiltonian, in which the Hubbard parameter $U$ denotes the on-site Coulomb repulsion and $\langle n_{i,\sigma}\rangle$ is the mean occupation of spin $\sigma$ at site $i$. For the tight-binding calculations we use $U = 0$ eV and for the MFH calculations we use $U = 4$ eV. Eigenenergies $\varepsilon_{n,\sigma}$ and state vectors $c_{n,\sigma}$ of the eigenstates $\psi_{n,\sigma}$ were obtained by numerical diagonalization of $\hat{H}$ in the basis of atomic orbitals centered at carbon sites $\vec{r}_i = (x_i, y_i, z_i = 0)$. The resulting molecular orbitals were then calculated using

$$\psi_{n,\sigma}(\vec{r}) = \sum_i c_{n,\sigma,i}\phi_{2p_z}(\vec{r} - \vec{r}_i),$$

using the Slater-type $2p_z$ orbital of carbon ($a_B$ - Bohr radius, $C_{eff} = 3.25$ - effective nuclear charge):

$$\phi_{2p_z}(\vec{r}) = \sqrt{\frac{C_{eff}^5}{32\pi a_B^5}} z e^{-C_{eff}\frac{|\vec{r}|}{2a_B}}.$$

The LDOS maps were generated by calculating the sum of the absolute square of the respective orbitals $\psi_{n,\sigma}$ and plot the resulting data at a fixed height of $z = 15$ Å above the molecular plane.

**DFT and CASSCF calculations**

Multiconfigurational quantum chemistry calculations were done with the CASSCF/SC-NEVPT2 method to obtain the first excitation energies of the DBT dimers (**2$_k$**, **3$_k$** and **4$_k$**) and the ferromagnetic trimer, employing the ORCA 5.0.4 and 6.0.1 packages, respectively. First, for the dimers, we performed a geometrical relaxation using the RI-MP2 method with the def2-SVP basis and the RIJCOSX approximation of the Coulomb and Hartree-Fock exchange with def2/J and def2-SVP/C auxiliary basis. The resulting natural orbitals and optimized geometry were used as input for the CASSCF calculation. Because of the high computational load of RI-MP2 when calculating the ferromagnetic trimer, we performed its relaxation with DFT using the B3LYP density functional (with the RIJCOSX approximation and the same basis set and auxiliary basis as the dimers), and the Kohn-Sham orbitals as input for CASSCF. As a constraint in both dimers and trimers, the C-C bond that connects the DBT units was not allowed to rotate, that is, the dihedral angle formed by the adjacent carbon atoms was fixed to zero.

For the dimers, a complete active space of 8 electrons and 12 orbitals was selected, that is, CASSCF(8,12), where the natural orbital with the largest fractional occupation, after optimization, always had above 1.70 electrons and the natural orbital with the smallest fractional occupation always had below 0.05 electrons. We performed the state average for the spin quantum numbers $S$ = 3, 2, 1 and 0 and a number of roots of 14, 18, 18 and 18, respectively, for the $2_k$ molecule; 14 roots for the four chosen multiplicities for $3_k$; and 6, 14, 14 and 14 roots for $4_k$. For the trimer, a complete active space of 8 electrons and 10 orbitals was selected, that is, CASSCF(8,10). The state average in this case was done for $S$ = 3, 2, 1 and 0 and 6 roots for each multiplicity. We found that four (six) of the optimized orbitals presented an occupation close to 1 electron in the case of the dimers (trimer). These electrons belong to the non-bonding orbitals (that is, zero/close to zero-energy modes in the tight-binding model) and match the expected multiradical character of the DBT dimers and trimer. After convergence, an additional strongly contracted NEVPT2 step was done to calculate first excitation energies with enhanced accuracy.

The spectral function displayed in Fig. 5 for the ferromagnetic trimer was calculated through an exact diagonalization of the CAS-Hubbard model. The active space consisted of 6 electrons in 6 orbitals (CAS(6,6)), with a nearest-neighbor hopping parameter $t$ = -2.7 eV and an on-site Coulomb repulsion $U = |t|$.

## Acknowledgments

R.O. would like to thank Karol Strutynski for fruitful discussions. This work was supported financially by the European Research Council Synergy grant MolDAM (grant no. 951519). Furthermore, support within the scope of the project CICECO-Aveiro Institute of Materials, UIDB/50011/2020 (DOI 10.54499/UIDB/50011/2020), UIDP/50011/2020 (DOI 10.54499/UIDP/50011/2020) & LA/P/0006/2020 (DOI 10.54499/LA/P/0006/2020), financed by national funds through the FCT/MCTES (PIDDAC) is gratefully acknowledged. This project has received funding from the European Union's Horizon 2020 research and innovation program, under grant agreement No 958174 and 101046231, and from the Foundation for Science and Technology (FCT) under grant agreement M-ERA-NET3/0006/2021 through the M-ERA.NET 2021 call. Financial support from the Spanish Agencia Estatal de Investigación (PID2022-140845OB-C62), the Xunta de Galicia (Centro de Investigación do Sistema Universitario de Galicia, 2023-2027, ED431G 2023/03) and the European Union (European Regional Development Fund – ERDF), is gratefully acknowledged.

## Author contributions

S.M., D.P. and L.G. conceived the experiment. M.V.V. synthesized and characterized the precursor in solution. F.P. and S.M. performed the on-surface synthesis and scanning probe experiments. F.P. performed tight-binding and mean-field Hubbard calculations. R.O. and M.M.-F. performed density functional theory and many-body calculations. F.P. and S.M. wrote the manuscript, and all authors contributed to discussing the results and revising the manuscript.

## Notes

The authors declare no competing financial interest.

## Supporting Information Available

Solution synthesis and characterization of **1** (Figures S1–S3), and scanning probe measurements and calculations on DBT dimers and trimers (Figures S4–S8).

## References

(1) Oteyza, D. G. de; Frederiksen, T. Carbon-Based Nanostructures as a Versatile Platform for Tunable π-Magnetism. *J. Phys.: Condens. Matter* **2022**, *34* (44), 443001. https://doi.org/10.1088/1361-648X/ac8a7f.

(2) Ovchinnikov, A. A. Multiplicity of the Ground State of Large Alternant Organic Molecules with Conjugated Bonds. *Theoret. Chim. Acta* **1978**, *47* (4), 297–304. https://doi.org/10.1007/BF00549259.

(3) Arikawa, S.; Shimizu, A.; Shiomi, D.; Sato, K.; Shintani, R. Synthesis and Isolation of a Kinetically Stabilized Crystalline Triangulene. *J. Am. Chem. Soc.* **2021**, *143* (46), 19599–19605. https://doi.org/10.1021/jacs.1c10151.

(4) Valenta, L.; Mayländer, M.; Kappeler, P.; Blacque, O.; Šolomek, T.; Richert, S.; Juríček, M. Trimesityltriangulene: A Persistent Derivative of Clar's Hydrocarbon. *Chem. Commun.* **2022**, *58* (18), 3019–3022. https://doi.org/10.1039/D2CC00352J.

(5) Pavliček, N.; Mistry, A.; Majzik, Z.; Moll, N.; Meyer, G.; Fox, D. J.; Gross, L. Synthesis and Characterization of Triangulene. *Nat. Nanotechnol.* **2017**, *12* (4), 308–311. https://doi.org/10.1038/nnano.2016.305.

(6) Mishra, S.; Beyer, D.; Eimre, K.; Liu, J.; Berger, R.; Gröning, O.; Pignedoli, C. A.; Müllen, K.; Fasel, R.; Feng, X.; Ruffieux, P. Synthesis and Characterization of π-Extended Triangulene. *J. Am. Chem. Soc.* **2019**, *141* (27), 10621–10625. https://doi.org/10.1021/jacs.9b05319.

(7) Su, J.; Telychko, M.; Hu, P.; Macam, G.; Mutombo, P.; Zhang, H.; Bao, Y.; Cheng, F.; Huang, Z.-Q.; Qiu, Z.; Tan, S. J. R.; Lin, H.; Jelínek, P.; Chuang, F.-C.; Wu, J.; Lu, J. Atomically Precise Bottom-up Synthesis of π-Extended [5]Triangulene. *Sci. Adv.* **2019**, *5* (7), eaav7717. https://doi.org/10.1126/sciadv.aav7717.

(8) Mishra, S.; Xu, K.; Eimre, K.; Komber, H.; Ma, J.; Pignedoli, C. A.; Fasel, R.; Feng, X.; Ruffieux, P. Synthesis and Characterization of [7]Triangulene. *Nanoscale* **2021**, *13* (3), 1624–1628. https://doi.org/10.1039/D0NR08181G.

(9) Turco, E.; Bernhardt, A.; Krane, N.; Valenta, L.; Fasel, R.; Juríček, M.; Ruffieux, P. Observation of the Magnetic Ground State of the Two Smallest Triangular Nanographenes. *JACS Au* **2023**, *3* (5), 1358–1364. https://doi.org/10.1021/jacsau.2c00666.

(10) Jacob, D.; Fernández-Rossier, J. Theory of Intermolecular Exchange in Coupled Spin-1/2 Nanographenes. *Phys. Rev. B* **2022**, *106* (20), 205405. https://doi.org/10.1103/PhysRevB.106.205405.

(11) Mishra, S.; Beyer, D.; Eimre, K.; Ortiz, R.; Fernández-Rossier, J.; Berger, R.; Gröning, O.; Pignedoli, C. A.; Fasel, R.; Feng, X.; Ruffieux, P. Collective All-Carbon Magnetism in Triangulene Dimers. *Angew. Chem. Int. Ed.* **2020**, *59* (29), 12041–12047. https://doi.org/10.1002/anie.202002687.

(12) Mishra, S.; Catarina, G.; Wu, F.; Ortiz, R.; Jacob, D.; Eimre, K.; Ma, J.; Pignedoli, C. A.; Feng, X.; Ruffieux, P.; Fernández-Rossier, J.; Fasel, R. Observation of Fractional Edge Excitations in Nanographene Spin Chains. *Nature* **2021**, *598* (7880), 287–292. https://doi.org/10.1038/s41586-021-03842-3.

(13) Zhao, C.; Catarina, G.; Zhang, J.-J.; Henriques, J. C. G.; Yang, L.; Ma, J.; Feng, X.; Gröning, O.; Ruffieux, P.; Fernández-Rossier, J.; Fasel, R. Tunable Topological Phases in Nanographene-Based Spin-1/2 Alternating-Exchange Heisenberg Chains. *Nat. Nanotechnol.* **2024**, 1–7. https://doi.org/10.1038/s41565-024-01805-z.

(14) Zhao, C.; Yang, L.; Henriques, J. C. G.; Ferri-Cortés, M.; Catarina, G.; Pignedoli, C. A.; Ma, J.; Feng, X.; Ruffieux, P.; Fernández-Rossier, J.; Fasel, R. Gapless Spin Excitations in Nanographene-Based Antiferromagnetic Spin-1/2 Heisenberg Chains. **2024**, arXiv:2408.10045 [cond-mat.mtrl-sci]. arXiv.org e-Print archive. https://doi.org/10.48550/arXiv.2408.10045 (accessed October 30, 2024).

(15) Su, X.; Ding, Z.; Hong, Y.; Ke, N.; Yan, K.; Li, C.; Jiang, Y.; Yu, P. Fabrication of Spin-1/2 Heisenberg Antiferromagnetic Chains via Combined On-Surface Synthesis and Reduction for Spinon Detection. **2024**, arxiv:2408.08801 [cond-mat.mes-hall]. arXiv.org e-Print archive. https://doi.org/10.48550/arXiv.2408.08801 (accessed October 30, 2024).

(16) Yuan, Z.; Zhang, X.-Y.; Jiang, Y.; Qian, X.; Wang, Y.; Liu, Y.; Liu, L.; Liu, X.; Guan, D.; Li, Y.; Zheng, H.; Liu, C.; Jia, J.; Qin, M.; Liu, P.-N.; Li, D.-Y.; Wang, S. Atomic-Scale Imaging of Fractional Spinon Quasiparticles in Open-Shell Triangulene Spin-1/2 Chains. **2024**, arxiv:2408.08612 [cond-mat.mtrl-sci]. arXiv.org e-Print archive. https://doi.org/10.48550/arXiv.2408.08612 (accessed October 30, 2024).

(17) Zheng, Y.; Li, C.; Xu, C.; Beyer, D.; Yue, X.; Zhao, Y.; Wang, G.; Guan, D.; Li, Y.; Zheng, H.; Liu, C.; Liu, J.; Wang, X.; Luo, W.; Feng, X.; Wang, S.; Jia, J. Designer Spin Order in Diradical Nanographenes. *Nat. Commun.* **2020**, *11* (1), 6076. https://doi.org/10.1038/s41467-020-19834-2.

(18) Cheng, S.; Xue, Z.; Li, C.; Liu, Y.; Xiang, L.; Ke, Y.; Yan, K.; Wang, S.; Yu, P. On-Surface Synthesis of Triangulene Trimers via Dehydration Reaction. *Nat. Commun.* **2022**, *13* (1), 1705. https://doi.org/10.1038/s41467-022-29371-9.

(19) Du, Q.; Su, X.; Liu, Y.; Jiang, Y.; Li, C.; Yan, K.; Ortiz, R.; Frederiksen, T.; Wang, S.; Yu, P. Orbital-Symmetry Effects on Magnetic Exchange in Open-Shell Nanographenes. *Nat. Commun.* **2023**, *14* (1), 4802. https://doi.org/10.1038/s41467-023-40542-0.

(20) Turco, E.; Wu, F.; Catarina, G.; Krane, N.; Ma, J.; Fasel, R.; Feng, X.; Ruffieux, P. Magnetic Excitations in Ferromagnetically Coupled Spin-1 Nanographenes. *Angew. Chem. Int. Ed.* **2024**, e202412353. https://doi.org/10.1002/anie.202412353.

(21) Daugherty, M. C.; Jacobse, P. H.; Jiang, J.; Jornet-Somoza, J.; Dorit, R.; Wang, Z.; Lu, J.; McCurdy, R.; Tang, W.; Rubio, A.; Louie, S. G.; Crommie, M. F.; Fischer, F. R. Regioselective On-Surface Synthesis of [3]Triangulene Graphene Nanoribbons. *J. Am. Chem. Soc.* **2024**, *146* (23), 15879–15886. https://doi.org/10.1021/jacs.4c02386.

(22) Zhu, X.; Li, K.; Liu, J.; Wang, Z.; Ding, Z.; Su, Y.; Yang, B.; Yan, K.; Li, G.; Yu, P. Topological Structure Realized in Cove-Edged Graphene Nanoribbons via Incorporation of Periodic Pentagon Rings. *J. Am. Chem. Soc.* **2024**, *146* (11), 7152–7158. https://doi.org/10.1021/jacs.4c00270.

(23) Mishra, S.; Beyer, D.; Eimre, K.; Kezilebieke, S.; Berger, R.; Gröning, O.; Pignedoli, C. A.; Müllen, K.; Liljeroth, P.; Ruffieux, P.; Feng, X.; Fasel, R. Topological Frustration Induces Unconventional Magnetism in a Nanographene. *Nat. Nanotechnol.* **2020**, *15* (1), 22–28. https://doi.org/10.1038/s41565-019-0577-9.

(24) Fernández-Rossier, J. Theory of Single-Spin Inelastic Tunneling Spectroscopy. *Phys. Rev. Lett.* **2009**, *102* (25), 256802. https://doi.org/10.1103/PhysRevLett.102.256802.

(25) Henriques, J. C. G.; Fernández-Rossier, J. Anatomy of Linear and Nonlinear Intermolecular Exchange in *S* = 1 Nanographene. *Phys. Rev. B* **2023**, *108* (15), 155423. https://doi.org/10.1103/PhysRevB.108.155423.

(26) Jacob, D.; Ortiz, R.; Fernández-Rossier, J. Renormalization of Spin Excitations and Kondo Effect in Open-Shell Nanographenes. *Phys. Rev. B* **2021**, *104* (7), 075404. https://doi.org/10.1103/PhysRevB.104.075404.

(27) de la Torre, B.; Švec, M.; Foti, G.; Krejčí, O.; Hapala, P.; Garcia-Lekue, A.; Frederiksen, T.; Zbořil, R.; Arnau, A.; Vázquez, H.; Jelínek, P. Submolecular Resolution by Variation of the Inelastic Electron Tunneling Spectroscopy Amplitude and Its Relation to the AFM/STM Signal. *Phys. Rev. Lett.* **2017**, *119* (16), 166001. https://doi.org/10.1103/PhysRevLett.119.166001.

(28) Meyer, G. A Simple Low-temperature Ultrahigh-vacuum Scanning Tunneling Microscope Capable of Atomic Manipulation. *Rev. Sci. Instrum.* **1996**, *67* (8), 2960–2965. https://doi.org/10.1063/1.1147080.

(29) Giessibl, F. J. High-Speed Force Sensor for Force Microscopy and Profilometry Utilizing a Quartz Tuning Fork. *Appl. Phys. Lett.* **1998**, *73* (26), 3956–3958. https://doi.org/10.1063/1.122948.

(30) Albrecht, T. R.; Grütter, P.; Horne, D.; Rugar, D. Frequency Modulation Detection Using High-Q Cantilevers for Enhanced Force Microscope Sensitivity. *J. Appl. Phys.* **1991**, *69* (2), 668–673. https://doi.org/10.1063/1.347347.

(31) Bartels, L.; Meyer, G.; Rieder, K.-H. Controlled Vertical Manipulation of Single CO Molecules with the Scanning Tunneling Microscope: A Route to Chemical Contrast. *Appl. Phys. Lett.* **1997**, *71* (2), 213–215. https://doi.org/10.1063/1.119503.

(32) Gross, L.; Mohn, F.; Moll, N.; Liljeroth, P.; Meyer, G. The Chemical Structure of a Molecule Resolved by Atomic Force Microscopy. *Science* **2009**, *325* (5944), 1110–1114. https://doi.org/10.1126/science.1176210.

# Supporting Information

## A Route Toward the On-Surface Synthesis of Organic Ferromagnetic Quantum Spin Chains

Fabian Paschke[1,*], Ricardo Ortiz[2], Shantanu Mishra[1,*], Manuel Vilas-Varela[3], Florian Albrecht[1], Diego Peña[3,4,*], Manuel Melle-Franco[2], Leo Gross[1,*]

[1]IBM Research Europe – Zurich, 8803 Rüschlikon, Switzerland
[2]CICECO - Institute of Materials, Department of Chemistry, University of Aveiro, Aveiro 3810-193, Portugal
[3]Center for Research in Biological Chemistry and Molecular Materials (CiQUS) and Department of Organic Chemistry, University of Santiago de Compostela, 15702 Santiago de Compostela, Spain
[4]Oportunius, Galician Innovation Agency (GAIN), 15702 Santiago de Compostela, Spain

[*]Corresponding authors: FAP@zurich.ibm.com, SHM@zurich.ibm.com, diego.pena@usc.es and LGR@zurich.ibm.com

**Contents:**

## Solution synthesis of molecular precursors

Starting materials were purchased reagent grade from TCI and Sigma-Aldrich and used without further purification. Pentacenequinone **1a** (Fig. S1) was obtained following a reported procedure.[1] Reactions were carried out in flame-dried glassware and under an inert atmosphere of purified Ar using Schlenk techniques. Thin-layer chromatography was performed on silica gel 60 F-254 plates (Merck). Column chromatography was performed on silica gel (40-60 µm). NMR spectra were recorded on a Bruker Varian Mercury 300 or Bruker Varian Inova 500 spectrometers. Mass spectra were recorded on a Bruker Micro-TOF spectrometer.

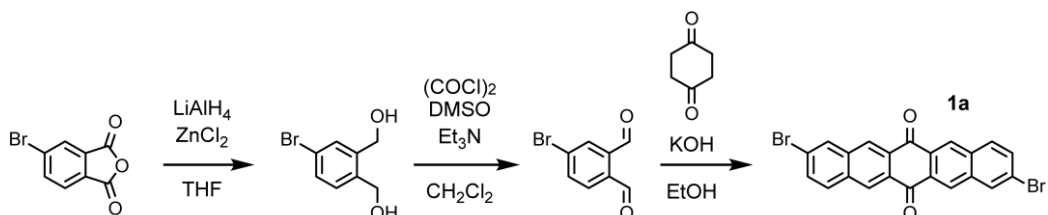

**Figure S1.** Synthesis of pentacenequinone **1a**.

Over a solution of **1a** (75 mg, 0.16 mmol) in $H_2SO_4$ (50 mL) at 50 °C, Al powder (300 mg, 11.1 mmol) was added, and the resulting mixture was stirred at 50 °C for 16 h (Fig. S2). Then, the reaction mixture was poured on ice and extracted with $CH_2Cl_2$ (2 × 125 mL) and EtOAc (125 mL). The combined organic extracts were dried over anhydrous $Na_2SO_4$, filtered, and evaporated under reduced pressure to obtain the crude pentacene derivative **1b** (40 mg).

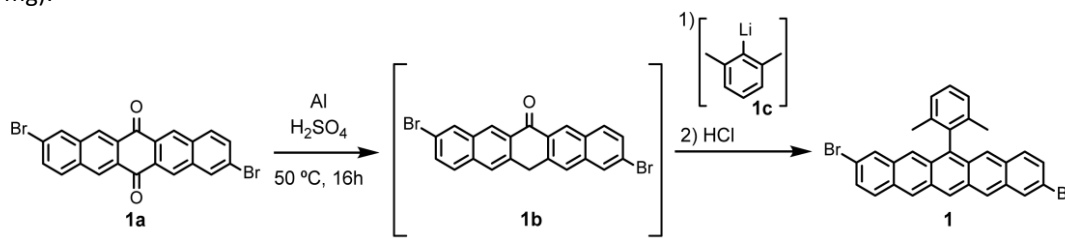

**Figure S2.** Synthesis of substituted **1**.

Compound **1b** (80 mg, 0.18 mmol) was suspended in anhydrous toluene (30 mL) and heated to 60 °C. Then, a solution of **1c** (4.8 mL, 0.15 M in $Et_2O$) was quickly added and the resulting mixture was stirred for 30 min. Then, $HCl_{(aq)}$ (2.00 mL, 37% wt) was added, and the resulting mixture was stirred at 60 °C for 5 min. The phases were separated, and the organic layer was evaporated under reduced pressure. The residue was purified by column chromatography (SiO₂; hexane:$CH_2Cl_2$ 4:1) affording **1** (1 mg, 1%) as a blue solid. ¹H-NMR (Fig. S3, 500 MHz, CDCl₃) δ: 9.01 (s, 1H), 8.70 (s, 1H), 8.63 (s, 1H), 8.13 (s, 1H), 8.04 (d, *J* = 5.6 Hz, 1H), 7.97 (d, *J* = 8.2 Hz, 2H), 7.82 (d, *J* = 9.2 Hz, 1H), 7.62 (d, *J* = 9.5 Hz, 1H), 7.48 (t, *J* = 7.5 Hz, 1H), 7.40 – 7.27 (m, 4H), 1.73 (s, 6H) ppm. MS (MALDI-TOF) for $C_{30}H_{20}Br_2$; calculated: 537.99, found: 538.00.

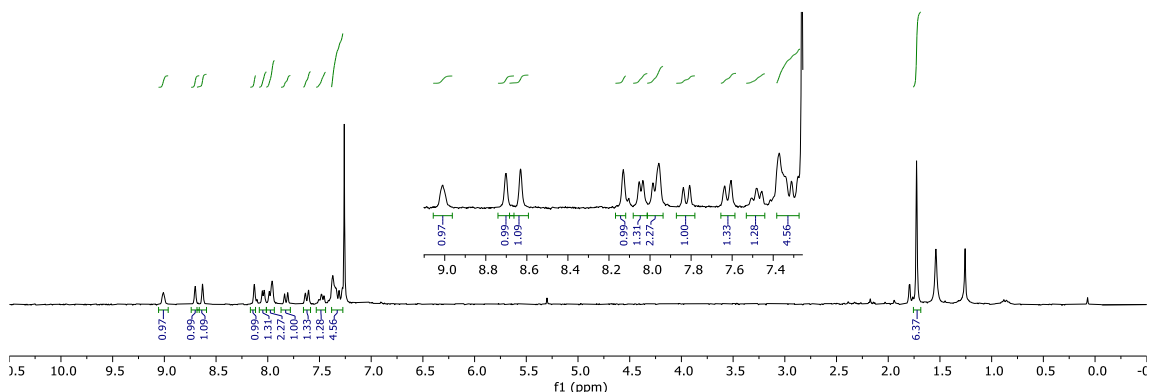

**Figure S3.** ¹H-NMR (500 MHz, CDCl₃) spectrum of **1**.

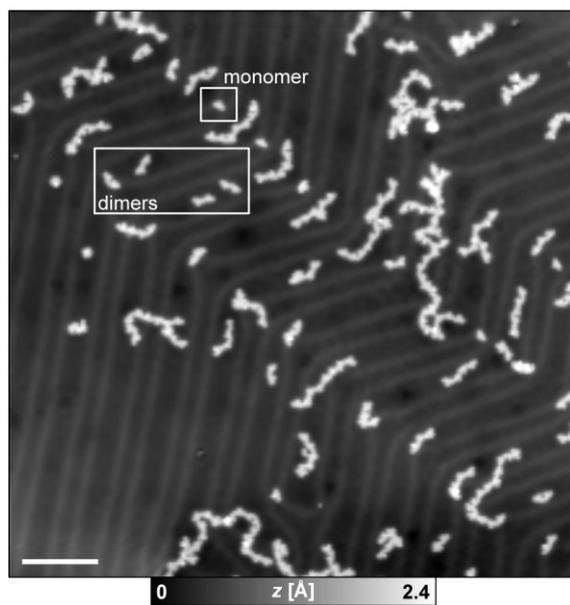

**Figure S4. On-surface synthesis of DBT monomers and oligomers from 1.** Overview STM image after deposition of **1** on a Au(111) surface and subsequent annealing steps (see Methods for details). We observed a mixture of individual molecules, dimers and longer oligomers on the surface. STM parameters: $V$ = 50 mV, $I$ = 20 pA. Scale bar: 10 nm.

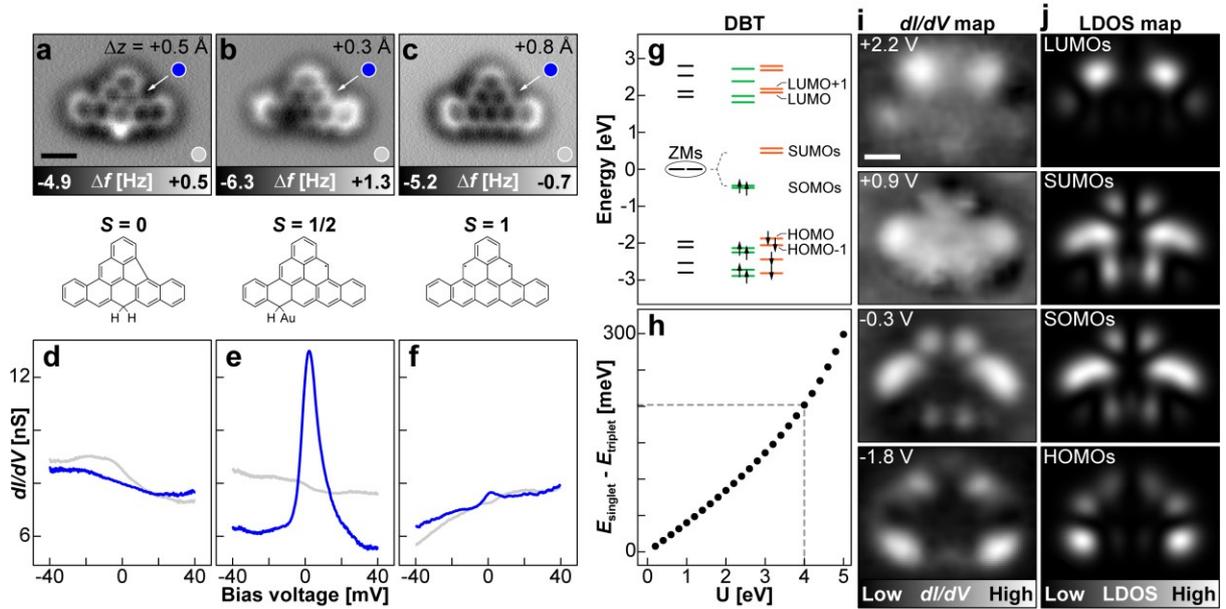

**Figure S5. Structural, electronic and magnetic properties of DBT. (a–c)** AFM images of defective (a, b) and intact (c) DBT monomers on Au(111). The corresponding chemical structures are drawn below each panel, with the respective spin ground states indicated. Note that the species in panel b was found pinned to the herringbone elbow site of Au(111), and the chemical structure is proposed based on ref.[2] **(d–f)** $dI/dV(V)$ spectra obtained on the species shown in panels a (d), b (e) and c (f), respectively. Arrows in panels a–c indicate the positions at which the blue spectra have been recorded. Background spectra on Au(111) are shown in grey. The absence of any low-energy spectral feature in panel d agrees with a closed-shell ground state of the species in panel a. A pronounced zero-bias peak in panel e is consistent with a $S = 1/2$ Kondo-screened ground state of the species in panel b. A comparatively weaker zero-bias peak in panel f agrees with an $S = 1$ underscreened Kondo state of DBT.[3] STM open-feedback parameters: $V$ = +40 mV, $I$ = 300 pA; $V_{rms}$ = 3.5 mV. **(g)** Tight-binding (black markers) and MFH (colored markers) ground-state energy spectra of DBT. Green (orange) markers denote spin up (down) levels. The MFH solution predicts a triplet ground state of DBT, in agreement with Ovchinnikov's rule. **(h)** MFH singlet-triplet gap of DBT, plotted as a function of $U$. The vertical dashed line indicates the value of $U$ (4 eV or $1.4t_1$) that is used throughout this article for MFH calculations. The singlet-triplet gap of DBT at $U$ = 4 eV is ~200 meV (horizontal dashed line), indicating a strong intra-DBT ferromagnetic exchange. **(i)** $dI/dV$ maps of the DBT molecule in panel c with the bias voltages indicated in the respective panels. STM parameters: $I$ = 300 pA; $V_{rms}$ = 50 mV. **(j)** MFH-LDOS maps of HOMOs (corresponding to equal contributions of HOMO and HOMO-1), SOMOs, SUMOs and LUMOs (corresponding to equal contributions of LUMO and LUMO+1) of DBT. Scale bars: 0.5 nm (applies to panels a–c, and all figures in panels i and j).

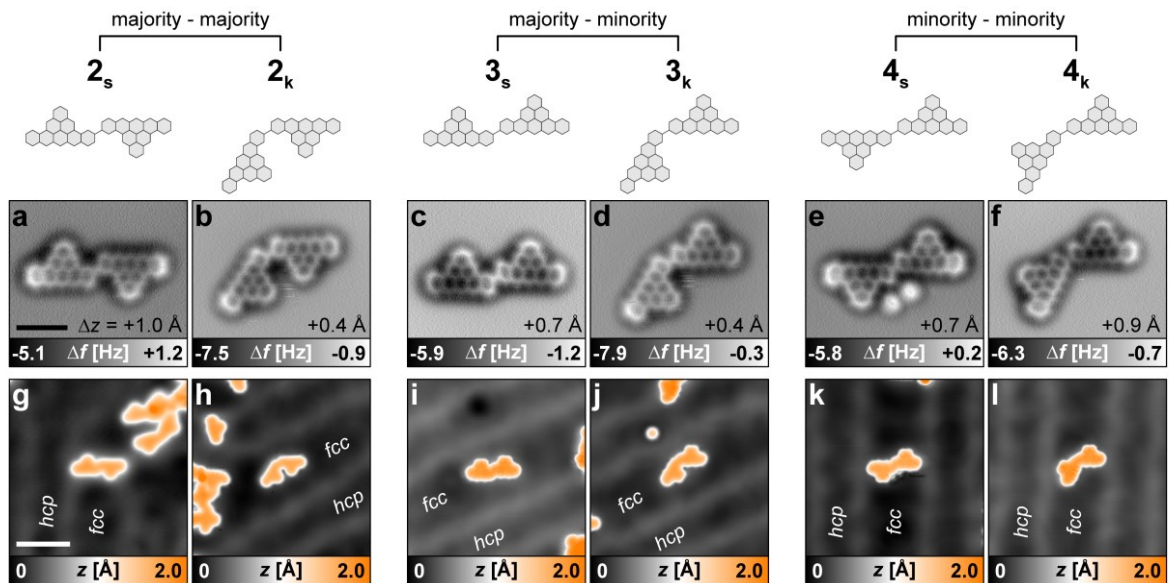

**Figure S6. STM and AFM imaging of kinked and straight DBT dimers. (a–f)** AFM images of kinked and straight DBT dimers (corresponding schematic representations are shown above the respective panels). Kinked dimers $2_k$, $3_k$ and $4_k$ are shown in Fig. 2 and discussed in the main text. STM and STS measurements on straight dimers $2_s$, $3_s$ and $4_s$ are shown in Fig. S7. The two bright spots adjacent to the dimer in panel e are co-adsorbed CO molecules. Out of 26 DBT dimers, 6, 9 and 11 exhibited majority-majority, majority-minority and minority-minority couplings, respectively. **(g–l)** Corresponding STM images of the dimers shown in panels a–f. Face-centered cubic (*fcc*) and hexagonal close-packed (*hcp*) regions of the reconstructed Au(111) surface are indicated. STM parameters: $V$ = +10 mV, $I$ = 10 pA. Scale bars: 1 nm (panels a–f) and 3 nm (panels g–l).

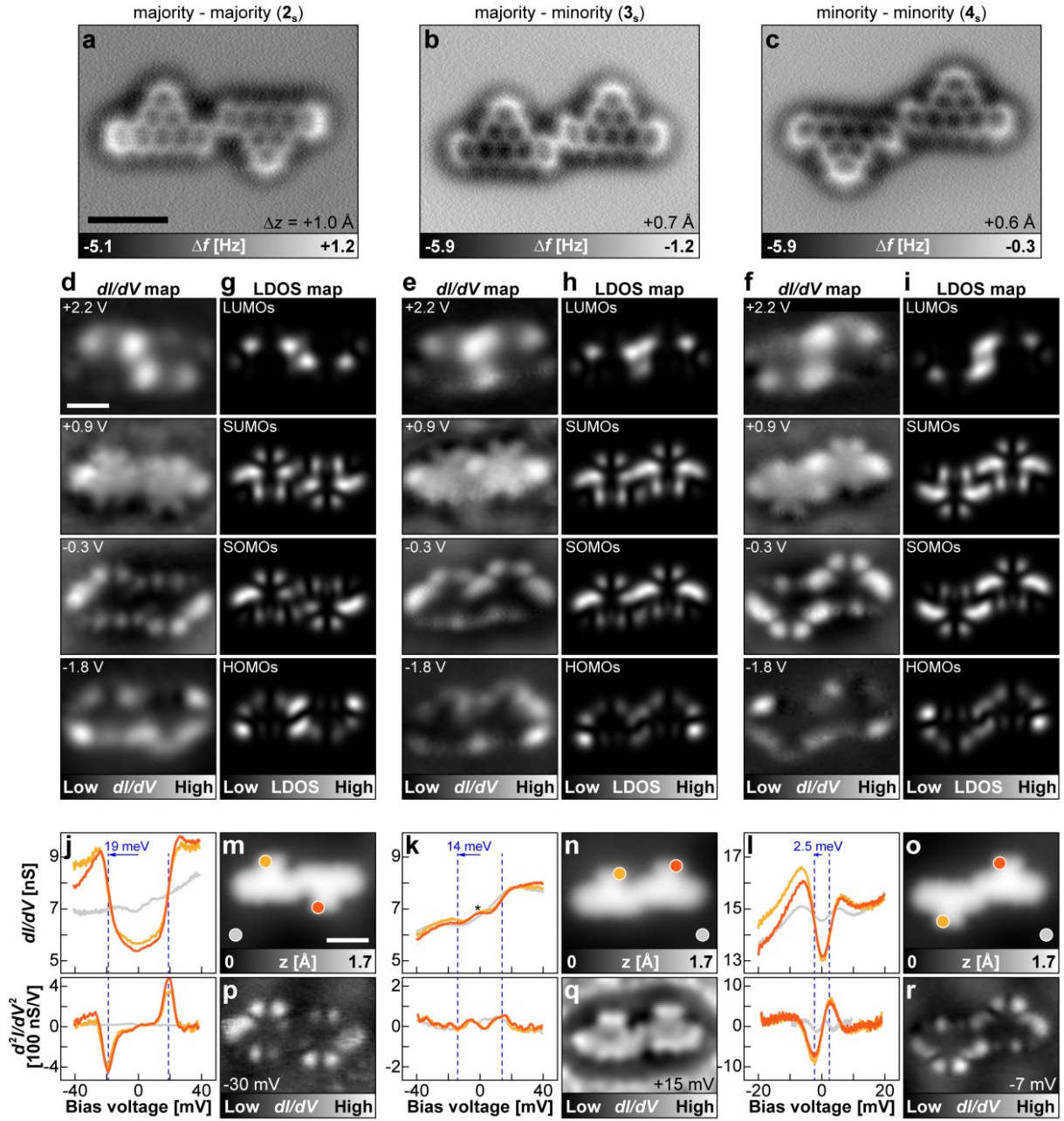

**Figure S7. Structural, electronic and magnetic characterization of straight DBT dimers. (a–c)** AFM images of $2_s$ (a), $3_s$ (b) and $4_s$ (c). **(d–f)** d$I$/d$V$ maps of $2_s$ (d), $3_s$ (e) and $4_s$ (f). Bias voltages at which the maps were acquired are indicated in the respective panels. STM parameters: $I$ = 300 pA; $V_{rms}$ = 50 mV. **(g–i)** Corresponding MFH-LDOS maps of the HOMOs to LUMOs of $2_s$ (g), $3_s$ (h) and $4_s$ (i). Orbital contributions for each map are the same as for $2_k$, $3_k$ and $4_k$ (see Fig. 3). **(j–l)** d$I$/d$V$($V$) (top) and d$^2I$/d$V^2$($V$) (bottom) spectra obtained on $2_s$ (j), $3_s$ (k) and $4_s$ (l). Background spectra on Au(111) are shown in grey. Dashed lines denote extracted spin excitation energies. The asterisk in panel k denotes the $S$ = 2 Kondo resonance of $3_s$. STM open-feedback parameters: $V$ = +40 mV, $I$ = 300 pA; $V_{rms}$ = 3.5 mV. **(m–o)** STM images of $2_s$ (m), $3_s$ (n) and $4_s$ (o). Colored circles indicate the positions at which spectra in panels j–l have been recorded. STM parameters: $V$ = +40 mV, $I$ = 300 pA. **(p–r)** Corresponding d$I$/d$V$ maps of the DBT dimers obtained around the respective spin excitation thresholds as indicated. STM parameters: $I$ = 300 pA; $V_{rms}$ = 3.5 mV. Scale bars: 1 nm (applies to panels a–c, and all figures in panels d–i and m–r).

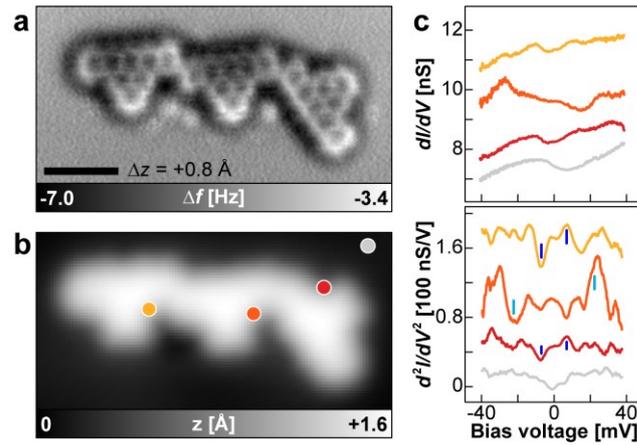

**Figure S8. Magnetic characterization of a ferromagnetic DBT trimer. (a)** AFM image of **5$_{FM}$** as shown in Fig. 5. **(b)** STM image of **5$_{FM}$**, obtained with a metallic (Au) tip. Colored circles indicate positions at which the spectra shown in panel c have been recorded. STM parameters: *V* = +40 mV, *I* = 300 pA. **(c)** *dI/dV*(*V*) (top) and *d²I/dV²*(*V*) (bottom) spectra obtained on the DBT trimer shown in panel a using a metallic (Au) tip, where no CO vibrational features are present (*c.f.* Fig. 5). A background spectrum obtained on Au(111) is shown in grey. Blue lines denote extracted spin excitation energies. For clarity, *dI/dV* (*d²I/dV²*) spectra on left, central and right units are vertically shifted by 4 nS (160 nS/V), 2 nS (100 nS/V) and 1 nS (40 nS/V), respectively. STM open-feedback parameters: *V* = +40 mV, *I* = 300 pA; *V*$_{rms}$ = 3.5 mV. Scale bar: 1 nm, applies to panel a and b.

**References:**

(1) Zhang, J.; Chen, Z.; Yang, L.; Pan, F.-F.; Yu, G.-A.; Yin, J.; Liu, S. H. Elaborately Tuning Intramolecular Electron Transfer Through Varying Oligoacene Linkers in the Bis(Diarylamino) Systems. *Sci. Rep.* **2016**, *6* (1), 36310. https://doi.org/10.1038/srep36310.

(2) Mishra, S.; Beyer, D.; Eimre, K.; Kezilebieke, S.; Berger, R.; Gröning, O.; Pignedoli, C. A.; Müllen, K.; Liljeroth, P.; Ruffieux, P.; Feng, X.; Fasel, R. Topological Frustration Induces Unconventional Magnetism in a Nanographene. *Nat. Nanotechnol.* **2020**, *15* (1), 22–28. https://doi.org/10.1038/s41565-019-0577-9.

(3) Turco, E.; Bernhardt, A.; Krane, N.; Valenta, L.; Fasel, R.; Juríček, M.; Ruffieux, P. Observation of the Magnetic Ground State of the Two Smallest Triangular Nanographenes. *JACS Au* **2023**, *3* (5), 1358–1364. https://doi.org/10.1021/jacsau.2c00666.